\documentclass[article,preprint,groupedaddress,11pt]{revtex4}
\usepackage{epsfig}
\usepackage{graphicx}
\usepackage{amssymb}

\begin{document}
\title{Fermat's principle in black-hole spacetimes}
\author{Shahar Hod}
\affiliation{The Ruppin Academic Center, Emeq Hefer 40250, Israel}
\affiliation{ }
\affiliation{The Hadassah Institute, Jerusalem 91010, Israel}
\date{\today}
\centerline {\it This essay received an Honorable Mention in the 2018 Essay Competition of the Gravity Research
Foundation}

\begin{abstract}
\ \ \ Black-hole spacetimes are known to possess closed light rings.
We here present a remarkably compact theorem which reveals the
physically intriguing fact that these unique null circular geodesics
provide the {\it fastest} way, as measured by asymptotic observers,
to circle around spinning Kerr black holes.
\newline
\newline
Email: shaharhod@gmail.com
\end{abstract}
\bigskip
\maketitle

Fermat's principle, also known as the principle of least time
\cite{Lip,Notewik}, asserts that among all possible null
trajectories, the path taken by a ray of light between two given
points A and B in a flat spacetime geometry is the path that {\it
minimizes} the traveling time $T_{A\to B}$. This remarkably elegant
principle implies, in particular, that the unique null trajectory
taken by a ray of light between two given points is generally
distinct from the straight line trajectory which minimizes the
spatial distance $d_{AB}$ between these points.

In the present Essay we would like to highlight an intriguing and
closely related physical phenomenon which characterizes curved
spacetime geometries. In particular, we here raise the physically
interesting question: Among all possible closed paths that circle
around a black hole in a curved spacetime, which path provides the
{\it fastest} way, as measured by asymptotic observers, to circle
the central black hole?

We first note that, in flat spacetimes, the characteristic orbital
period $T_{\odot}$ of a test particle that moves around a spatially
compact object of radius $R$ is trivially bounded from below by the
compact relation \cite{Noteunit}
\begin{equation}\label{Eq1}
T_{\odot}\geq T^{\text{flat}}_{\odot\text{min}}=2\pi R\  ,
\end{equation}
where the equality sign in (\ref{Eq1}) is attained by massless
particles that circle the compact object on the shortest possible
(tangential) trajectory with $r_{\text{fast}}=R$.

It should be emphasized, however, that the simple lower bound
(\ref{Eq1}) is not valid in realistic curved spacetimes. In
particular, it does not take into account the important time
dilation (red-shift) effect which is caused by the gravitational
field of the central compact object \cite{Chan}. In addition, the
flat-space relation (\ref{Eq1}) does not take into account the
well-known phenomenon of dragging of inertial frames by spinning
compact objects in curved spacetimes \cite{Chan}.

As we shall explicitly show below, due to the influences of these
two interesting physical effects, the shortest possible orbital
period $T_{\odot}$ of a test particle around a central compact
object, as measured by asymptotic observers, is larger than the
naive flat-space estimate (\ref{Eq1}). In particular, we shall prove
that, in generic curved spacetimes, the unique circular trajectory
$r=r_{\text{fast}}$ that minimizes the traveling time $T_{\odot}$
around a central Kerr black hole is distinct from the tangential
trajectory with $r=r_{\text{short}}$ which could minimize the
traveling distance around the spinning black hole.

{\it The fastest circular orbit around a spinning Kerr black
hole.---} We shall analyze the physical and mathematical properties
of equatorial circular trajectories around spinning Kerr black
holes. In Boyer-Lindquist coordinates $(t,r,\theta,\phi)$, the
asymptotically flat black-hole spacetime can be described by the
curved line element \cite{Chan}
\begin{eqnarray}\label{Eq2}
ds^2=-{{\Delta}\over{\rho^2}}(dt-a\sin^2\theta
d\phi)^2+{{\rho^2}\over{\Delta}}dr^2+\rho^2
d\theta^2+{{\sin^2\theta}\over{\rho^2}}\big[a
dt-(r^2+a^2)d\phi\big]^2\  ,
\end{eqnarray}
where $M$ is the black-hole mass, $J\equiv Ma$ is its angular
momentum, $\Delta\equiv r^2-2Mr+a^2$, and $\rho^2\equiv
r^2+a^2\cos^2\theta$. The black-hole (event and inner) horizons are
determined by the spatial zeros of the metric function $\Delta$:
\begin{equation}\label{Eq3}
r_{\pm}=M\pm(M^2-a^2)^{1/2}\  .
\end{equation}

We would like to identify the unique circular trajectory which
minimizes the orbital period $T_{\odot}$, as measured by asymptotic
observers, around the central black hole. We shall therefore assume
the relation $v/c\to 1^-$ for the velocity of the orbiting test
particle \cite{Notegng}. The corresponding radius-dependent orbital
periods $T_{\odot}(r)$ of the test particles can easily be obtained
from the characteristic black-hole curved line element (\ref{Eq2})
with $ds=dr=d\theta=0$ and $\Delta\phi=\pm2\pi$
\cite{Notecocoun,Noteco,Hod1}. This yields the compact functional
relation
\begin{equation}\label{Eq4}
T_{\odot}(r)=2\pi\cdot{{\sqrt{r^2-2Mr+a^2}-{{2Ma}\over{r}}}\over{1-{{2M}\over{r}}}}\
\end{equation}
for the orbital periods of co-rotating test particles around the
central spinning black hole.

The physically interesting co-rotating circular orbit with
$r=r_{\text{fast}}$, which is characterized by the shortest possible
orbital period $T_{\odot\text{min}}=\min_r\{T_{\odot}(r)\}$ around
the central spinning black hole, is determined by the functional
relation $dT(r=r_{\text{fast}})/dr=0$. This yields the
characteristic algebraic equation
\begin{equation}\label{Eq5}
r^2-3Mr+2a^2+2a\sqrt{r^2-2Mr+a^2}=0\ \ \ \ \text{for}\ \ \ \
r=r_{\text{fast}}\  .
\end{equation}
Remarkably, this equation can be solved analytically to yield the
simple functional relation
\begin{equation}\label{Eq6}
r_{\text{fast}}=2M\cdot\big\{1+\cos[{2\over
3}\cos^{-1}(-{a/M})]\big\}\ .
\end{equation}
for the unique orbital radius $r_{\text{fast}}(M,a)$ which
characterizes the fastest co-rotating circular trajectory (the
closed circular path with the shortest possible orbital period)
around the central spinning Kerr black hole.

What we find most intriguing is the fact that the spin-dependent
radii $r_{\text{fast}}(M,a)$ of the fastest circular trajectories,
as given by the functional expression (\ref{Eq6}), exactly coincide
with the corresponding radii $r_{\gamma}(M,a)$ of the null circular
geodesics \cite{Bar} which characterize the spinning Kerr black-hole
spacetimes. One therefore concludes that co-rotating null circular
geodesics (closed light rings) provide the fastest way, as measured
by asymptotic observers, to circle around generic Kerr black holes.

It is physically interesting to define the dimensionless ratio [see
Eqs. (\ref{Eq1}) and (\ref{Eq3})]
\begin{equation}\label{Eq7}
\Theta({\bar a})\equiv {{T_{\odot\text{min}}}\over{2\pi r_+}}\ \ \ \
; \ \ \ \ {\bar a}\equiv a/M\  ,
\end{equation}
which characterizes the unique closed circular trajectories [with
$r=r_{\text{fast}}({\bar a})$] that minimize the orbital periods
around the central spinning black holes. As emphasized above, a
naive flat-space calculation predicts the relation
$\Theta^{\text{flat}}_{\text{min}}\equiv
T^{\text{flat}}_{\odot\text{min}}/2\pi R=1$ [see Eq. (\ref{Eq1})].
However, substituting Eqs. (\ref{Eq4}) and (\ref{Eq6}) into
(\ref{Eq7}), one finds the characteristic inequality
\cite{Notetheta}
\begin{equation}\label{Eq8}
\Theta^{\text{Kerr}}({\bar a})>1\
\end{equation}
for all Kerr black-hole spacetimes in the physically allowed regime
${\bar a}\in[0,1]$. In particular, the dimensionless function
$\Theta^{\text{Kerr}}({\bar a})$ exhibits a non-trivial
(non-monotonic) functional dependence on the dimensionless
black-hole rotation parameter ${\bar a}$ with the property
\cite{Noteexpan}
\begin{equation}\label{Eq9}
\text{min}_{\bar a}\big\{\Theta^{\text{Kerr}}({\bar a})\big\}\simeq
2-{{3(13-7\sqrt{3})}\over{88}}\ \ \ \ \text{at}\ \ \ \ {\bar
a}^{\text{Kerr}}_{\text{min}}\simeq 1-{{126-45\sqrt{3}}\over{1936}}\
.
\end{equation}

{\it Summary.---} Fermat's principle asserts that, in a flat
spacetime geometry, the path taken by a ray of light is unique in
the sense that it represents the spatial trajectory with the
shortest possible traveling time between two given points
\cite{Lip,Notewik}. This intriguing principle implies, in
particular, that the paths taken by light rays are generally
distinct from the straight line trajectories which could minimize
the traveling distances between two given points.

In the present short Essay we have highlighted an intriguing and
closely related phenomenon in curved black-hole spacetimes. In
particular, we have raised the physically interesting question:
Among all possible trajectories that circle around a spinning Kerr
black hole, which closed trajectory provides the {\it fastest} way,
as measured by asymptotic observers, to circle the central black
hole?

Our compact theorem has revealed the physically intriguing fact that
the equatorial null circular geodesics (closed light rings), which
characterize the curved black-hole spacetimes, provide the fastest
way to circle around spinning Kerr black holes. In particular, we
have explicitly proved that, in analogy with the Fermat principle in
flat spacetime geometries, the unique curved trajectories
$r=r_{\text{fast}}(M,a)$ [see Eq. (\ref{Eq6})] which minimize the
traveling times $T_{\odot}$ of test particles around central black
holes are distinct from the tangential trajectories $r=r_+(M,a)$
[see Eq. (\ref{Eq3})] which could minimize the traveling distances
around the black holes.

\newpage

\noindent
{\bf ACKNOWLEDGMENTS}
\bigskip

This research is supported by the Carmel Science Foundation. I would
like to thank Yael Oren, Arbel M. Ongo, Ayelet B. Lata, and Alona B.
Tea for helpful discussions.



\begin{thebibliography}{99}
\bibitem{Noteunit} We use natural units in which $G=c=1$.
\bibitem{Notegng} In the present essay we consider geodesic as well
as non-geodesic trajectories of test particles around the central
Kerr black holes. It is well known that the null circular geodesic
of the black-hole spacetime, on which massless particles that move
exactly at the speed of light can circle the central black hole, is
characterized by a well defined radius $r_{\gamma}=r_{\gamma}(M,a)$.
It is important to stress the fact that, by using man-made rockets
which are based on non-gravitational forces, massive particles can
also circle the central black hole on non-geodesic trajectories with
orbital velocities that, in principle, may approach arbitrarily
close to the speed of light and with orbital radii that may differ
from the unique radius $r_{\gamma}(M,a)$ of the black-hole null
circular geodesic.
\bibitem{Noteexpan} Here we have used the series expansion $\Theta^{\text{Kerr}}({\bar
a})=2+\sqrt{2}(\sqrt{3}-2)\cdot\sqrt{\epsilon}+(14/3-2\sqrt{3})\cdot\epsilon+O(\epsilon^{3/2})$
in the near-extremal $0\leq \epsilon\equiv 1-{\bar a}\ll1$ regime.
\end{thebibliography}
\end{document}